**Atomistic modeling of interfacial segregation and structural transitions in ternary alloys**

Yang Hu [1], Timothy J. Rupert [1,2,*]
[1] Materials Science and Engineering, University of California, Irvine, California 92697, USA
[2] Mechanical and Aerospace Engineering, University of California, Irvine, California 92697, USA
* To whom correspondence should be addressed: trupert@uci.edu

Grain boundary engineering via dopant segregation can dramatically change the properties of a material. For metallic systems, most current studies concerning interfacial segregation and subsequent transitions of grain boundary structure are limited to binary alloys, yet many important alloy systems contain more than one type of dopant. In this work, hybrid Monte Carlo/molecular dynamics simulations are performed to investigate the behavior of dopants at interfaces in two model ternary alloy systems: Cu-Zr-Ag and Al-Zr-Cu. Trends in boundary segregation are studied, as well as the propensity for the grain boundary structure to become disordered at high temperature and doping concentration. For Al-Zr-Cu, we find that the two solutes prefer to occupy different sites at the grain boundary, leading to a synergistic doping effect. Alternatively, for Cu-Zr-Ag, there is site competition because the preferred segregation sites are the same. Finally, we find that thicker amorphous intergranular films can be formed in ternary systems by controlling the concentration ratio of different solute elements.

# 1. Introduction

Grain boundaries can influence a variety of material properties such as tensile strength, fracture toughness, fatigue resistance, ductility, formability, creep resistance, densification rate during sintering, and thermal stability [1-6]. For instance, grain boundaries can help increase the rate of shrinkage during the intermediate and final stages of sintering [7-10]. As vacancy sinks, grain boundaries absorb the vacancies which come from the surface of pores. These vacancies diffuse and accumulate along grain boundaries and then eventually collapse, which contributes to densification [7-10]. Another example would be the importance of grain boundaries for nanocrystalline metals, which contain a large volume of boundaries and have very high strength yet low ductility and thermal stability [11-16]. The increased strength of a nanocrystalline metal comes from the hindrance of dislocation motion by the grain boundaries, meaning higher stresses are needed to further deform the material [11-13]. However, another consequence of the limited motion of dislocations is that local stress concentrations at boundaries can lead to crack or voids formation near the grain boundaries and thus reduces the strain-to-failure of the material [11-13]. Moreover, grain boundaries have higher energies than perfect crystals, which provides a driving force for grain growth [14-16]. Therefore, nanocrystalline metals are usually not thermally stable and once the grains grow, they will lose their advantage of high strength.

In general, grain boundaries can have both positive and negative effects on material properties. Fortunately, some of these negative effects can be reduced by "grain boundary engineering" or the alteration of grain boundary character and/or the interconnecting grain boundary network. One possible method of grain boundary engineering is the use of solute segregation, which allows grain boundary structure and properties to be adjusted [17, 18]. For example, successful attempts at "segregation engineering" have occurred for MgO or NiO-doped alumina, where reduced grain boundary mobility and higher densification was found during the final stage of sintering [19, 20]. Grain boundary segregation can also be used to stabilize nanoscale grain sizes, because segregated dopants reduce both grain boundary energy and mobility [14-16]. For example, stable nanostructures were found in alloy systems such as Cu-Zr [21-23], Cu-Hf [21], Ni-Zr [21], W-Ti [24], Cu-Bi [25], and Cu-Cr [26].

Besides the local chemistry change, another product of dopant segregation to grain boundary sites can be a concomitant alteration of boundary structure. One way to describe such changes is through the use of the terminology "grain boundary complexions," which describes an interfacial state that cannot exist without adjacent grains but is able to undergo phase-like transitions [27-29]. Dillon et al. [27] reported on six different types of complexions in alumina, using thickness as their main criteria. These authors identified structures that included "clean" grain boundaries, a monolayer segregation pattern, bilayer segregation, multi-layer segregation, nanoscale intergranular films, and wetting films. Structurally disordered versions of the last two types can be called amorphous intergranular films (AIFs) and have been shown to have positive effects in some cases. For example, Khalajhedayati et al. [22, 23] managed to produce AIFs with stable nanoscale thicknesses in nanocrystalline Cu-Zr alloys, with in situ bending experiments showing a much higher strain-to-failure for the samples with AIFs than those with ordered grain boundaries. Using atomistic simulations, Pan and Rupert [30] provided an explanation for the improved ductility of nanostructured Cu-Zr with AIFs, finding that AIFs can efficiently absorb incoming dislocations during plastic deformation and reduce the stress concentration at grain boundaries, which in turn delays crack initiation and propagation. These authors also showed that the samples with thicker AIFs have better performance, indicating that thicker films gave better ductility. AIFs have also be shown to be helpful for materials processing, by encouraging rapid sintering of powder materials [31]. Activated sintering well below the bulk eutectic temperature has been observed in systems such as $Bi_2O_3$-doped ZnO [32], Ni-doped W [33], and CuO-doped $TiO_2$ [34]. For all of these materials, "liquid-like" intergranular films, another term used to describe an AIF, were found to form. The excess free volume in these "liquid-like" films lead to enhanced mass transport and a faster sintering rate [32-34].

The work of Pan and Rupert [30] that was already discussed demonstrates the power of atomistic modeling for simulating segregation-induced interfacial phenomenon and investigating the physical mechanisms behind experimental observations, but many more applications of these techniques can be found in the literature. For instance, the segregation of Ta and formation of Ta-rich clusters at grain boundaries in ultra-fine grained Cu-Ta was reported both experimentally and theoretically [35-37], with the Zener pinning effect of the Ta clusters explained by the simulations. To model the final segregated structure in these Cu-Ta alloys, a hybrid Monte Carlo/molecular dynamics (MC/MD) method [38, 39] was used by Koju et al. [37] to let the computational cell achieve both structural and chemical equilibrium much faster than pure MD simulations. A structural transition of grain boundaries to AIFs in Cu-Ag was also reported using MC/MD simulations, with Ag segregation driving the transition [40, 41]. The opposite process, Cu segregation at to grain boundaries and twin boundary defects in Ag, has been observed by Ke and Sansoz [42] in nanotwinned Ag with less than 1 at.% of Cu. Enhanced twin stability and yield strength were found and correlated with Cu segregation.

While progress has been made on understanding segregation and complexion transitions at the atomic scale, all of the studies mentioned above were limited to binary alloys. While such studies demonstrate the principle of segregation-induced grain boundary transitions, there are a number of cases where understanding ternary or higher element systems would be beneficial. First, "tramp elements" can often enter materials during processing and will complicate grain boundary segregation behavior. An example of this is high energy ball milling, where Fe atoms can be incorporated into the targeted alloy from the hardened steel milling media and vial [43]. In addition, one may actually plan to have a complex chemical composition at the boundary. For example, AIFs were discussed previously and have been observed in binary alloys, yet these features fundamentally exist because the local boundary composition lowers the energy penalty for the formation of an amorphous film [44, 45]. The literature on bulk metallic glasses has shown that three or more elements are needed to make good glass formers [46], suggesting that a grain boundary composition of three or more elements might lead to thicker AIFs. Finally, experimental reports in the literature show that more chemically complex alloys can be very interesting. As an example, recent impressive work on so called "high-entropy complexions" in Ni-based alloys have shown that multi-element segregation may lead to increased grain boundary configurational entropy that helps to reduce the grain boundary energy and stabilize a nanocrystalline microstructure [47]. As a whole, ternary and higher element systems are intriguing but complex, meaning that more study is required.

In this paper, we perform hybrid MC/MD simulations of interfacial segregation and structural transitions to investigate these behaviors with atomic-level detail. Two ternary alloy systems, Cu-Zr-Ag and Al-Zr-Cu, are used as model material systems, with the hope of discovering general rules that are applicable to other multi-component alloy systems. We find that the dopants compete with each other during segregation if they have the same preferred segregation sites, while others can have a synergistic effect which enhances the extent of segregation. Beyond simple dopant segregation, we also explore whether thicker AIFs can be stabilized by the addition of a third element to the grain boundary region, with the ratio of dopant concentrations found to be very important.

## 2. Methods

A hybrid MC/MD method was used to simulate interfacial segregation and structural transformations, using the implementation of Sadigh et al. [48]. Classical MD simulations allow for structure relaxation and were performed using the Large-scale Atomic/Molecular Massively Parallel Simulator (LAMMPS) package with an integration time step of 1 fs [49]. An isothermal-isobaric (NPT) ensemble with the Nose-Hoover thermostat/barostat was applied to relax samples at different temperatures (350 K and 600 K for the Al-based alloys, 600 K and 1000 K for the Cu-based alloys) and under zero pressure. 600 K and 1000 K are considered as high temperatures, based on a comparison with the melting temperatures ($T_m$) of pure Al and Cu obtained by the interatomic potentials ($T_m = 1175$ K for Cu and $T_m = 785$ K for Al, meaning 600 K is $0.76T_m$ for Al and 1000 K is $0.85\ T_m$ for Cu). Dopant concentrations in the range of 0-10 at.% were explored in the various binary and ternary alloys of interest. While high temperature and high dopant boundary concentration should promote the formation of AIFs (if the alloy can sustain them at all), ordered grain boundaries have been reported at low temperatures and dopant concentrations [23]. To allow for chemical relaxation, MC simulations in a variance-constrained semi-grand canonical ensemble were performed after every 100 MD steps. The system is considered to have found an equilibrium state when the absolute value of the fitted slope of the potential energy over the prior 1000 MC steps is less than $0.001$ eV step$^{-1}$. Finally, a conjugate gradient energy minimization was performed to remove thermal noise, so that the interfacial structure obtained during the doping process can be preserved and only the interfacial regions are identified as defects. Similar conjugate gradient energy minimization calculations were also employed to explore the segregation energies of solutes at different positions in the grain boundary at 0 K.

Bicrystal models with two $\Sigma 5$ (310) grain boundaries were used as starting configurations. The $\Sigma 5$ (310) grain boundary can be thought of as a model of a high-angle and high-energy tilt grain boundary in face centered cubic (fcc) metals. This boundary has an energy that is similar to a general high-angle grain boundary [50, 51] and a variety of potential doping sites [30], but still has a compact, repeating kite-shape structure that allows for efficient simulation, as shown in Fig. 1a. Different positions in the kite-shape structure are marked using numbers in this figure. The Σ5 (310) clean grain boundary at 0 K in Al is asymmetric, as shown in Fig. 1b, with the repeating structural unit being two kites connected to each other by sharing site 3 and 4 from the symmetric structure. This asymmetric structure is truly more stable than the symmetric version, as the calculated 0 K grain boundary energy of the asymmetric structure (460.63 mJ/m$^2$) is lower than the symmetric structure (464.75 mJ/m$^2$). However, the symmetric structure becomes the low energy structure at finite temperature, as shown in Fig. 1c, which means all of our simulations on interfacial segregation and structural transition occur at symmetric grain boundaries. The simulation box has a length of approximately 23 nm (X direction), a height of 11 nm (Y direction), and a thickness of 4 nm (Z direction), containing a total of 95,520 atoms. Additional samples that were longer in the X direction (~69 nm for Cu-based alloys) were used for calculating segregation energies of solutes, to reduce the interaction between the grain boundary and the solutes in the grain interior. All structural analysis and visualization of atomic configurations was performed using the open-source visualization tool OVITO [52] with Cu atoms shown as orange, Zr atoms shown as blue, Ag atoms shown as red, and Al atoms as pink. The local crystal structure of each atom was identified according to common neighbor analysis (CNA) [53], with fcc atoms colored green, hexagonal close packed atoms colored red, body centered cubic atoms colored blue, icosahedral atoms colored yellow, and "other" atoms color white.

The Cu-Zr-Ag and Al-Zr-Cu alloy systems were chosen for this study for a number of reasons. First, there is evidence that all of the added dopants should segregate to defects. Zr [21-23] and Ag [40, 41, 54] have been observed to segregate to grain boundaries in Cu-based alloys, while the interfacial segregation of Cu [55-57] in Al-based alloys has also been reported. Similarly, Tsivoulas and Robson [58] observed Zr enrichment along dislocations, followed by Al$_3$Zr precipitation in the as-cast and fully homogenised Al-Cu-Li based alloys. Second, Zr [21-23] and Ag [40, 41] addition by themselves have been reported to facilitate AIF formation in the Cu-Ag and Cu-Zr binary systems. Therefore, the Cu-Zr-Ag alloy is an obvious candidate to explore the effect of a shift to ternary chemical compositions. Although there are no direct studies about AIF formation in binary Al-Cu and Al-Zr, researchers have investigated the glass forming ability of binary metallic glasses with these chemical compositions, as well as the ternary Al-Zr-Cu system [59-61]. For example, Yan et al. [59] managed to obtain a stable amorphous phase in Al-Cu for a Cu concentration range of 26-36 at.%. While these two ternary systems are selected to observe both interfacial segregation and AIF formation, they do differ from each other when considering other aspects such as the atomic radius mismatch and enthalpy of mixing. For the Cu-Zr-Ag system, the atomic sizes of Zr and Ag are 25% and 12.5% larger than Cu, respectively. Atomic size mismatches of >12% have historically aided metallic glass formation [46]. For the Al-Zr-Cu system, the atomic size of Zr is 11.9% larger than Al while the atomic size of Cu is 10.5% smaller than Al. Cu-Zr, Al-Zr and Al-Cu all have negative enthalpies of mixing, while Cu-Ag has a positive enthalpy of mixing. A negative enthalpy of mixing is typically required for a good glass former in the metallic glass literature [46].

Two embedded-atom-method (EAM) potentials [62], originally used to model metallic glasses and generated by Sheng and coworkers, were used to describe the interactions between atoms [63, 64]. In our previous study about the identification of interatomic potentials for the accurate modeling of interfacial segregation and structural transitions [65], we found that the accurate reproduction of physical properties such as the enthalpy of mixing and bond energies are important for realistic modeling of such behavior. To validate the potentials used here, the liquid enthalpies of mixing for the four individual binary systems are calculated and shown in Fig. 2 (details of the calculation method can be found in Ref. [65]). The simulation temperatures for each alloy system are much higher than the melting temperature of Cu (1358

K) and Al (937 K) [66], and they are chosen so that our results could be compared with available experimental data which also use temperatures above melting [67-70]. Fig. 2 shows that the interatomic potentials are in reasonable agreement with experimental data from the literature. While the numbers are not always exact matches, particularly for the Cu-rich alloys, the overall shapes and general trends agree. Physical properties such as the bulk modulus, shear modulus, and surface energies of Cu, Al, Ag, and $Zr_{fcc}$ (Zr atoms in an fcc arrangement) are also calculated for the two potentials and compared with experimental data or ab-initio calculations in Table 1. Overall, the calculated values from the two interatomic potentials are in reasonable agreement with the data from the literature. In summary, these two ternary potentials faithfully reproduce important physical properties needed to model dopant segregation to interfaces. Both potentials were constructed using the force-matching method, meaning that atomic forces (the negative derivatives of the potential-energy surface) are included in the fitting database and thus are able to better capture the interaction between atoms.

## 3. Results and discussion
### 3.1 Segregation and structural transitions in binary alloys

The behavior of the binary systems were first studied to provide a baseline for comparison. Doped grain boundaries in the Cu-Zr, Cu-Ag, Al-Cu, and Al-Zr systems are shown in Fig. 3. In this figure, the chemical information is shown in the frames on the left of each figure part, while the structural information is presented in the frames on the right. Atomic snapshots of samples simulated at 350 K (Al-rich alloys) or 600 K (Cu-rich alloys) and with 0.4 at.% or 1 at.% dopant concentration are presented to display representative examples of ordered grain boundaries. Additional samples simulated at higher temperatures of 600 K (Al-rich alloys) or 1000 K (Cu-rich alloys) and with larger dopant concentrations are chosen to show representative examples of conditions that are conducive to AIF formation. Fig. 3 shows that interfacial segregation of dopants occurs in all four binary systems, but the structural transitions observed at the grain boundaries are quite different. In the Cu-Zr and Cu-Ag alloys with ordered grain boundaries, the segregating dopants prefer to occupy the kite-tip sites (site 1 in Fig. 1a) and form what can be classified as a monolayer complexion at the grain boundary. At high temperature and doping concentration, AIFs form in both of these alloys. The difference between the two is that AIFs in the Cu-Ag system contain a collection of large Ag-rich clusters, meaning the boundary is structurally disordered but chemically ordered. Because of the positive enthalpy of mixing, Ag atoms cluster at the grain boundary as the global dopant concentration increases and, once the Ag-rich clusters become large enough, connect to one another to form an amorphous film. This process is shown in more detail in Fig. 4, with the change in structure occurring as dopant concentration increases. Ag-rich clusters only form in the samples at intermediate doping concentrations, such as Cu-5 at.% Ag shown in Fig. 4b, which may cause non-uniformity in the thickness of AIFs. In addition, these clusters are found to be energetically favorable and stable even if our convergence criteria for MD/MC is made ten times more restrictive or if the sample is relaxed in an NPT ensemble for a much longer time. Similar evolution of the grain boundary structure was also observed by Li and Szlufarska [41], who studied polycrystalline Cu-Ag nano-alloys. Since these authors performed simulations at a low temperature (300 K), they observed precipitation of fcc-Ag at the grain boundaries after Ag concentration reached a critical level. The formation of a liquid-like film at the grain boundary is also consistent with the prior observations of Williams and Mishin [40]. In contrast, the disordered boundary in Cu-Zr (Fig. 3b) is both structurally and chemically disordered.

For Al-Cu with ordered grain boundaries (Fig. 3e), multiple grain boundary sites are occupied. Some Cu atoms are found at site 4 (the bottom of the kite unit), while others are found along the edges of the kite unit in site 2. Finally, some dopants become interstitials that occupy the empty space in the kite-shaped structure. An AIF that is both structurally and chemically disordered forms in the sample doped by 9 at.% Cu at 600 K. For the Al-Zr system (Figs. 3g and 3h), the grain boundary remains ordered even after the global Zr concentration reaches 10 at.% and the temperature reaches 600 K. As mentioned previously, researchers who observed Zr segregation to dislocations in Al-based alloys found $Al_3Zr$ precipitates at dislocations, not amorphous phases, meaning that Zr atoms prefer to stay in ordered, crystalline phases when added to Al [58]. Indeed, careful inspection of the Al-Zr samples reveals the formation of coherent $L1_2$-type $Al_3Zr$ precipitates in the grain interior. The lack of an AIF in Al-Zr suggests that the energy penalty for forming an amorphous complexions is simply too large to make it energetically favorable, even when heavily doped at high temperature. Figs. 3e and 3g confirm that the doped Al boundaries at finite temperature have a symmetric structure. Two possible factors that could lead to such a shift are: (1) a thermal effect due to nonzero temperature and (2) atom redistribution that occurs when adding dopants. The fact that the same symmetric structure is found in the pure Al sample (Fig. 1c) suggests that thermal effects alone are enough to cause this change. However, Yang et al. did find that doping can cause shifts between symmetric and asymmetric boundary structure in Mo-Ni [80], meaning that further stabilization may come from segregation in the alloys.

To understand the details of segregation, it is useful to calculate the segregation energy, $\Delta H_{seg}$, of a single dopant at different sites in the grain boundary using molecular statics. The segregation energy is defined as the difference in energy between a simulation cell with dopants at the grain boundary and a simulation cell with the same amount of dopants in the grain interior. The equation used for calculating the segregation energy is [81]:

$$\Delta H_{seg}(\sum N_i) = \frac{(H(\sum N_i) - H(0) - \sum N_i \Delta H_{i,bulk})}{\sum N_i} \quad (1)$$

where $\Delta H_{seg}(\Sigma N_i)$ is the segregation energy per solute for a bicrystalline sample containing $\Sigma N_i$

dopants, $H(\Sigma N_i)$ is the energy of a bicrystalline sample containing $\Sigma N_i$ dopants, $\Delta H_{i,bulk}$ is the change in energy when inserting one dopant, $i$, into the bulk. A negative value of segregation energy indicates a preference for segregation, while a positive value signals grain boundary depletion. The segregation energy as a function of distance from the interface plane is shown in Fig. 5, along with a view of the boundary looking down the (100) axis. The most promising or lowest energy segregation sites for each dopant are circled in green. For Cu-Zr and Cu-Ag, the most advantageous segregation site is site 1 (kite-tip). Since the segregation energy of Zr (-1.3 eV) is more negative than that of Ag (-0.53 eV) for site 1, we hypothesize that Zr will fill these sites first when we extend our analysis to the ternary Cu-based alloys. DFT calculations suggest that the segregation energy of Zr at the symmetric Σ5 (310) boundary is -1.62 eV [82], showing that the calculated values here are reasonable enough to look for trends. The next most promising site for Zr and Ag segregation is site 3 (two identical locations at the outside of the kite-bottom). For Al-based alloys, the symmetric grain boundary simulated by the ternary potential is not stable at 0 K, as previously mentioned. This means that the symmetric structure will transform and be removed under the energy minimizations used for molecular statics calculations. Therefore, to calculate the segregation energy of dopants in the Al-rich alloys, a symmetric grain boundary obtained by a binary Al-Cu potential [55] was used as the starting configuration. Energy minimization were subsequently performed on samples with dopants at different segregation sites for 100 minimization steps, which allowed the symmetric grain boundary shape to be retained. The absolute values of the numbers associated with the segregation energies for the Al-rich systems may therefore not be the true minimum values, but trends can still be extracted from Figures 5c and d. The preferred segregation sites for Zr are also the kite-tip sites, while Cu favors the kite-bottom site and two sites at the edge of the kite shape structure. Note that the sites marked with "X" labels in Figure 5d were not considered, as even our mild minimization treatment lead to a rearrangement to an asymmetric structure in this case.

Dopant segregation to a grain boundary can reduce the grain boundary energy, as well as remove any elastic strain energy penalty that is induced by substituting dopants into the lattice [83]. The dependence of interfacial segregation of dopants or other defects like vacancies on local distortion has been reported in a variety of material systems [84, 85]. To understand if there is a correlation between the solute segregation pattern and local stress here, the atomic hydrostatic stresses in pure Cu and pure Al in the vicinity of the grain boundaries was obtained and is presented in Fig. 6. The sites under large positive (i.e., tensile) hydrostatic stress at both boundaries are the sites with large negative segregation energies for Zr and Ag in Cu and Zr in Al (Fig. 5), while those sites under large compression are the favorable sites for dopants smaller than the matrix element, such as Cu atoms in Al. This agrees with the notion that atoms larger than the host element will prefer sites under tension while atoms smaller than the host element will prefer sites that are compressed. The alignment of the preferred sites according to the local hydrostatic stresses suggests that mechanical effects play a dominant role in determining segregation sites, while chemical effects may be of secondary importance in these alloys.

**3.2 Segregation and structural transitions in ternary alloys**

At low temperature and low dopant concentration, ordered grain boundaries were found in both Cu-Zr-Ag and Al-Zr-Cu. Chemical and structural snapshots of representative ordered grain boundaries, as well as their solute concentration profiles and the Gibbsian interfacial excesses are shown in Fig. 7. The Gibbsian interfacial excess of solutes describes the total solute excess per unit area and is defined as [86]:

$$\Gamma = \frac{N_{GB}(c_{GB} - c_0)}{A} \tag{2}$$

where $N_{GB}$ is the number of atoms in grain boundary region, $c_{GB}$ is the grain boundary composition, $c_0$ is the grain interior composition, and $A$ is the interfacial area. Fig. 7a shows that the ordered boundary in the Cu-Zr-Ag alloy is characterized by Zr atoms at the kite-tip sites (site 1), while no Ag atoms can be found at the sites that were most promising in the binary Cu-Ag. Instead, the Ag atoms must be incorporated as a solid solution into the lattice. Fig. 7c shows this behavior in a more quantitative manner, where a drop in the Ag concentration at the grain boundary is found while there is a sudden increase in the Zr concentration. The total dopant concentration (black triangles) is approximately equal to the Zr concentration (blue circles) in the grain boundary region at the center of Fig. 7c, but then it is equal to the Ag concentration (red diamonds) in the grain interior. Since Zr interfacial segregation leads to the formation of a monolayer complexion at the grain boundary, meaning that all the Zr atoms are located along one atomic plane, the increase of Zr concentration is very sharp and is limited to a small region. Fig. 7e presents the Gibbsian interfacial excess in the ternary Cu-based alloys and shows a similar trend, with the curve for Ag having a small negative value at the interface. Taken as a whole, these figures show that, although Zr and Ag would like to segregate to the same promising segregation sites, Zr wins out and fills them first. This observation is consistent with our calculation of a lower segregation energy in absolute terms for Zr in Cu, as compared to Ag in Cu. Although there is another potential segregation site for Zr and Ag at the two sides of the bottom of the kite (site 3 in Fig. 1), this site is rarely occupied. This likely occurs because the local strain fields from having two oversized dopants near each other would lead to an overall energy increase. The segregation energy of Ag to site 3 was -0.16 eV in the binary Cu-Ag calculation, but this value drops to 0 eV after all of the kite-tip sites are occupied by Zr.

In contrast, Zr and Cu occupy different sites at grain boundaries in Al-Zr-Cu, as shown in Fig. 7b. From Fig. 7d and 7f, the concentrations of both dopant species peak in the grain boundary, signaling the co-segregation of Cu and Zr. Since the two types of dopants prefer different sites, they do not compete and in fact the grain boundary can be much more heavily doped. This can be seen by observing that the maximum grain boundary dopant concentration in Al-Zr-Cu is more than twice as large as that in the Cu-Zr-Ag alloy (~40 at.% versus ~17 at.%, respectively). While the global dopant concentrations are not the same, the value of ~17 at.% for Cu-Zr-Ag is the maximum that can be achieved with a monolayer segregation pattern for this simulation cell. This suggests that careful selection of dopant elements is important when planning for segregation engineering. For example, if one wants to maximize the grain boundary doping to reduce boundary energy and limit grain growth, a synergistic type of co-segregation should be targeted. At high temperature and dopant concentration, AIFs form in both ternary systems. Chemical and structural snapshots of representative amorphous grain boundary films, as well as their solute concentration profiles and the Gibbsian interfacial excesses are shown in Fig. 8. Both alloys show structurally and chemically disordered amorphous films at the grain boundary, but the AIF in Al-Zr-Cu contains only a small amount of Zr. Fig. 8c presents the total dopant concentration in Cu-2 at.% Zr-2 at.% Ag, where both Zr and Ag concentrations increase in the grain boundary region, indicating co-segregation. Unlike the competition for limited segregation sites in the ordered grain boundary which leads to the depletion of Ag, there are many more possible segregation sites generated after the formation of a disordered structure in the AIF. Because of this, Ag also has a chance to segregate to the grain boundaries. In the Al-5 at.% Zr-5 at.% Cu alloy shown in Fig. 8d, the concentration of Cu reaches its highest value in the grain boundary, but there is a drop in local Zr concentration. Since the two dopants had no trouble being close to each other in the ordered boundary, it is unlikely that the reduced Zr is caused by a chemical repulsion of the two dopants. A more likely explanation is that there is simply an energetic penalty associated with Zr in an amorphous structure here, which was why AIFs did not form in the binary Al-Zr alloy. The curves in Fig. 8e and 8f show the Gibbsian interfacial excess of dopants, which have similar shapes to the chemical concentration curves. In Fig. 8f, there is a range of negative Gibbsian interfacial excess for Zr at the grain boundary, again signaling depletion of Zr.

To understand the difference between segregation in binary and ternary systems, the Gibbsian interfacial excess is presented in Fig. 9 as a function of the global dopant concentration for the ternary systems, with the sum of the Gibbsian interfacial excess for the two binary systems included for comparison (i.e., the data point at 4 at.% for the binary alloys in this figure sums the contributions for Cu-2 at.% Zr and Cu-2 at.% Ag). The concentration ratio of different types of dopants in the ternary alloys is 1:1 for simplicity, with the simulations performed at 600 K for the Al-based alloys and 1000 K for the Cu-based alloys, which are considered high temperatures. In Fig. 9a, both curves increase as the total dopant concentration increases, intersecting at ~1.5 at.%. This value of ~1.5 at.% dopants is also the concentration where the ternary alloy first begins to form an AIF. When the boundary is ordered, segregation competition occurs and the grain boundary dopant concentration is reduced as a result. After AIF formation, there are many more potential doping sites and the grain boundary dopant concentration quickly moves past the binary systems. Gibbsian interfacial excess data is shown in Fig. 9b for the Al-based alloys where a similar trend is found, although the difference between the ternary and the two binary alloys is more obvious. In this case, the intersection of the two curves happens at ~7.5 at.%. The structural transition to AIFs occurs before this dopant concentration and the slope of the data noticeably increases above ~6 at.%. Both cases show that adding a second segregating element is beneficial for grain boundary segregation, once amorphous films are formed.

**3.3 Forming thicker amorphous intergranular films in ternary systems**

Pan and Rupert reported that AIFs can improve ductility by delaying crack nucleation and slowing down crack propagation, with this effect increasing as film thickness increases [30].

Thus, a question of great practical importance is whether a ternary alloy can sustain thicker AIFs. Fig. 10 presents the grain boundary thickness as a function of the total dopant concentration for both ternary and binary Cu-based and Al-based alloys. Again, we hold the concentration ratio of different types of dopants in the ternary systems as 1:1 for now, with all simulations performed at 600 K for Al-based alloys and 1000 K for Cu-based alloys. Fig. 10a and 10c show that the ternary curves lie between the curves for the two related binary systems. The Cu-Zr and Al-Cu systems have the thickest AIFs for a given total dopant concentration. Cu-Ag forms AIFs but they are relatively thin, which can be explained by the clustering of Ag caused by the positive enthalpy of mixing. Note that we excluded the data points of Cu-Ag samples in which there are Ag-rich clusters formed rather than uniform AIFs. A second caveat exists for the Cu-Ag system. The prior work showing that AIFs can delay crack nucleation and propagation focused on amorphous films which were both structurally and chemically disordered (i.e., like the films in Cu-Zr and Al-Cu). It is not obvious whether structurally disordered but chemically ordered films such as those in Cu-Ag would be beneficial for mechanical properties. Finally, Al-Zr cannot sustain AIFs, so the grain boundary thickness remains small. Interestingly, the curve for Al-Zr in Fig. 10c shows a modest decrease from the pure Al grain boundary thickness. In this case, the "thickness" only comes from thermal fluctuations. Zr segregates to the grain boundary in Al, reduces the amount of thermal fluctuation, and hence the grain boundary is thinner. Fig. 10b and 10d isolate the ternary alloys and compare them with the binary that is better able to sustain AIFs (Cu-Zr and Al-Cu), while keeping the amount of Zr and Cu constant. In this case, we are asking when AIFs are the thickest for a given amount of Zr or Cu in the system. This would be important if a given amount of one dopant is limited (e.g., if >X at.% Zr led to precipitation of an intermetallic phase). These figures show that the curves for ternary systems are above the curves for the two binary systems, showing that ternary compositions may still be useful in certain situations.

As a final test of the ability of these ternary alloys to form AIFs, we relax the restriction on the concentration ratio of the different dopants. It is possible that there is a promising concentration ratio that allows ternary systems to form thicker AIFs than the binaries for a given amount of total dopant, as the metallic glass literature is filled with examples of alloys where the best glass forming composition is not an even distribution between the constituent elements [61, 87-89]. To explore this, we fix the total amount of dopant to a given value while changing the concentration ratio of the two dopants in the ternary alloy. For the Cu-Zr-Ag system with a total concentration of 4 at.% dopant, film thickness is presented in Fig. 11a as a function of the Zr concentration. This means that the binary Cu-Ag appears on the plot at 0 at.% and the binary Cu-Zr appears on the plot at 4 at.%, with ternary alloys in between. For this material, we find that the thickness simply moves between the two bounds. The film thickness versus Cu concentration for Al-Zr-Cu with 10 at.% total dopant is shown in Fig. 11b, with 0 at.% indicating the Al-Zr binary system and 10 at.% indicating the Al-Cu binary system. In this case, we do find that a ternary mixture can more efficiently produce AIFs, as the thickest AIFs are found in for Al-9.5 at.% Zr-0.5 at.% Cu. While the ternary AIFs only appear to be slightly thicker in absolute numbers, it is important to put this number in perspective, as they are a measurable 4.5% thicker than the AIFs in Al-Cu. This observation proves that ternary compositions can in fact lead to thicker AIFs compared to binary alloys in certain situations. It is worth noting that we observe this behavior even though one of the two binaries was a relatively poor glass former, as Al-Zr did not amorphize at all. Future work should focus on combinations where all alloy combinations are good glass formers in their binary combinations.

## 4. Summary and Conclusions

This paper reports on atomistic simulations that were performed to study interfacial segregation and amorphous complexion formation in ternary alloys, with Cu-Zr-Ag and Al-Zr-Cu used as example systems. A bicrystal sample with two $\Sigma 5$ (310) grain boundaries is used as a model to represent a high-angle and high-energy tilt grain boundary. The following conclusions can be drawn from this work:

- In Cu-Zr-Ag, the preferred segregation site for both Zr and Ag is the kite-tip site, leading to segregation competition. This leads to a local depletion of Ag at the grain boundary in the ternary alloys with ordered grain boundaries. In contrast, for Al-Zr-Cu, Zr and Cu have different preferred segregation sites and they therefore have a synergistic doping pattern. In the end, this leads to much higher grain boundary dopant concentrations in Al-Zr-Cu.
- The segregation patterns in the ordered grain boundaries for these alloy systems can be explained by mechanical effects, where the local hydrostatic stress and relative atomic size determine site preference.
- By controlling the concentration ratio of different dopant elements, it is possible to form thicker AIFs in some ternary alloys systems. Alternatively, if only a limited amount of a given dopant species can be added, a ternary composition can be a way to enable further grain boundary segregation and thickening of AIFs.

Ternary alloys exhibit more complicated segregation behavior than their binary counterparts, but this complexity brings additional opportunities as well. For example, one motivating factor of our study was the observation that AIFs can toughen nanocrystalline metals. Since we have found that ternaries can experience added grain boundary segregation and thicker AIFs, ternary nanocrystalline materials should be studied in more detail.

**Acknowledgements**

This research was supported by U.S. Department of Energy, Office of Basic Energy Sciences, Materials Science and Engineering Division under Award No. DE-SC0014232.

**References**

[1] McLean D (1957) Grain boundaries in metals 1st edn, Clarendon Press, Oxford.
[2] Sutton AP and Balluffi RW (2006) Interfaces in crystalline materials, Oxford University Press, New York.
[3] Randle V (1993) The measurement of grain boundary geometry 1st edn, Taylor and Francis, London.
[4] Howe JM (1997) Interfaces in materials: atomic structure, thermodynamics and kinetics of solid-vapor, solid-liquid and solid-solid interfaces, Wiley-Interscience, New York.
[5] Gottstein G and Shvindlerman LS (2009) Grain boundary migration in metals: thermodynamics, kinetics, applications 2nd edn, CRC press, New York.
[6] Wolf D and Yip S (1992) Materials interfaces: atomic-level structure and properties 1st edn, CRC press, New York.
[7] Alexander BH, Balluffi RW (1957) The mechanism of sintering of copper, Acta Metall 5: 666-677.
[8] Burke JE (1957) Role of Grain Boundaries in Sintering, J Am Ceram Soc 40:80-85.
[9] Coble RL, Burke JE (1963) Sintering in Ceramics, Progr Ceram Sci 3:197-251
[10] Djohari H, Derby JJ (2009) Transport mechanisms and densification during sintering: II. Grain boundaries, Chem Eng Sci 64:3810-3816.
[11] Meyers MA, Mishra A, Benson DJ (2006) Mechanical properties of nanocrystalline materials, Prog Mater Sci 51:427-556.
[12] Kumar KS, Van Swygenhoven H, Suresh S (2003) Mechanical behavior of nanocrystalline metals and alloys1, Acta Mater 51:5743-74.
[13] Dao M, Lu L, Asaro RJ, De Hosson JT, Ma E (2007) Toward a quantitative understanding of mechanical behavior of nanocrystalline metals, Acta Mater 55:4041-65.
[14] Mathaudhu SN, Boyce BL (2015) Thermal stability: the next frontier for nanocrystalline materials, JOM 67:2785-7.
[15] Kalidindi AR, Chookajorn T, Schuh CA (2015) Nanocrystalline materials at equilibrium: a thermodynamic review, JOM 67:2834-43.
[16] Peng HR, Gong MM, Chen YZ, Liu F (2017) Thermal stability of nanocrystalline materials: thermodynamics and kinetics, Int Mater Rev 62:303-33.
[17] Raabe D, Herbig M, Sandlobes S, Li Y, Tytko D, Kuzmina M, Ponge D and Choi PP (2014) Grain boundary segregation engineering in metallic alloys: A pathway to the design of interfaces, Curr Opin Solid St M 18:253-61.
[18] Seah MP (1980) Grain-Boundary Segregation, J Phys F Met Phys 10:1043-64.
[19] Jorgensen PJ and Westbrook JH (1964) Role of Solute Segregation at Grain Boundaries During Final–Stage Sintering of Alumina, J Am Ceram Soc 47:332-338.
[20] Jorgensen PJ (1965) Modification of Sintering Kinetics by Solute Segregation in $Al_2O_3$, J Am Ceram Soc 48:207-210.
[21] Schuler JD, Rupert TJ (2017) Materials selection rules for amorphous complexion formation in binary metallic alloys, Acta Mater 140:196-205.
[22] Khalajhedayati A, Pan ZL, Rupert TJ (2016) Manipulating the interfacial structure of nanomaterials to achieve a unique combination of strength and ductility, Nat Commun 7:10802.
[23] Khalajhedayati A, Rupert TJ (2015) High-temperature stability and grain boundary complexion formation in a nanocrystalline Cu-Zr alloy, JOM 67:2788-801.
[24] Chookajorn T, Murdoch HA, Schuh CA (2012) Design of stable nanocrystalline alloys, Science 337:951-4.
[25] Mayr SG, Bedorf D (2007) Stabilization of Cu nanostructures by grain boundary doping with Bi: Experiment versus molecular dynamics simulation, Phys Rev B 76:024111.

[26] Harzer TP, Djaziri S, Raghavan R, Dehm G (2015) Nanostructure and mechanical behavior of metastable Cu–Cr thin films grown by molecular beam epitaxy, Acta Mater 83:318-32.
[27] Dillon SJ, Tang M, Carter WC, Harmer MP (2007) Complexion: A new concept for kinetic engineering in materials science, Acta. Mater 55:6208-6218.
[28] Harmer MP (2011) The Phase Behavior of Interfaces, Science 332:182-183.
[29] Cantwell PR, Tang M, Dillon SJ, Luo J, Rohrer GS, Harmer MP (2014) Grain boundary complexions, Acta Mater 62:1-48.
[30] Pan Z, Rupert TJ (2015) Amorphous intergranular films as toughening structural features, Acta Mater 89:205-14.
[31] Luo J (2008) Liquid-like interface complexion: From activated sintering to grain boundary diagrams, Curr Opin Solid State Mater Sci 12: 81-88.
[32] Luo J, Wang H and Chiang Y (1999) Origin of Solid- State Activated Sintering in $Bi_2O_3$-Doped ZnO, J Am Ceram Soc 82: 916-920.
[33] Gupta VK, Yoon DH, Meyer HM, Luo J (2007) Thin intergranular films and solid-state activated sintering in nickel-doped tungsten, Acta Mater 55:3131-3142.
[34] Nie J, Chan JM, Qin M, Zhou N, Luo J (2017) Liquid-like grain boundary complexion and sub-eutectic activated sintering in CuO-doped $TiO_2$, Acta Mater 130:329-338.
[35] Darling KA, Rajagopalan M, Komarasamy M, Bhatia MA, Hornbuckle BC, Mishra RS, Solanki KN (2016) Extreme creep resistance in a microstructurally stable nanocrystalline alloy, Nature 537:378-81.
[36] Rajagopalan M, Darling K, Turnage S, Koju RK, Hornbuckle B, Mishin Y, Solanki KN (2017) Microstructural evolution in a nanocrystalline Cu-Ta alloy: A combined in-situ TEM and atomistic study, Mater Des 113:178-85.
[37] Koju RK, Darling KA, Kecskes LJ, Mishin Y (2016) Zener pinning of grain boundaries and structural stability of immiscible alloys, JOM 68:1596-604.
[38] Mishin Y (2014) Calculation of the $\gamma/\gamma'$ interface free energy in the Ni–Al system by the capillary fluctuation method, Model Simul Mater Sci Eng 22:045001.
[39] Pun GP, Yamakov V, Mishin Y (2015) Interatomic potential for the ternary Ni–Al–Co system and application to atomistic modeling of the B2–L10 martensitic transformation, Model Simul Mater Sci Eng 23:065006.
[40] Williams PL, Mishin Y (2009) Thermodynamics of grain boundary premelting in alloys. II. Atomistic simulation, Acta Mater 57:3786-3794.
[41] Li A, Szlufarska I (2017) Morphology and mechanical properties of nanocrystalline Cu/Ag alloy, J Mater Sci, 52(8):4555-67.
[42] Ke X, Sansoz F (2017) Segregation-affected yielding and stability in nanotwinned silver by microalloying, Phys Rev Mater 1(6):063604.
[43] Cipolloni G, Pellizzari M, Molinari A, Hebda M, Zadra M (2015) Contamination during the high-energy milling of atomized copper powder and its effects on spark plasma sintering, Powder Technol 275:51-59.
[44] Zhou NX, Luo J (2015) Developing grain boundary diagrams for multicomponent alloys, Acta Mater 91:202-216.
[45] Zhou NX, Hu T, Luo J (2016) Grain boundary complexions in multicomponent alloys: Challenges and opportunities, Curr Opin Solid State Mater Sci 20:268-277.
[46] Inoue A (2000) Stabilization of metallic supercooled liquid and bulk amorphous alloys, Acta Mater 48:279-306.
[47] Zhou N, Hu T, Huang J, Luo J (2016) Stabilization of nanocrystalline alloys at high temperatures via utilizing high-entropy grain boundary complexions, Scripta Mater 124:160-163.

[48] Sadigh B, Erhart P, Stukowski A, Caro A, Martinez E, Zepeda-Ruiz L (2012) Scalable parallel Monte Carlo algorithm for atomistic simulations of precipitation in alloys, Phys Rev B 85:184203.
[49] Plimpton S (1995) Fast Parallel Algorithms for Short-Range Molecular-Dynamics, J Comput Phys 117:1-19.
[50] Zhang L, Lu C, Tieu K (2014) Atomistic simulation of tensile deformation behavior of Σ5 tilt grain boundaries in copper bicrystal, Sci Rep 4:5919.
[51] Tschopp MA, Coleman SP, McDowell DL (2015) Symmetric and asymmetric tilt grain boundary structure and energy in Cu and Al (and transferability to other fcc metals), Integrating Materials and Manufacturing Innovation 4:11.
[52] Stukowski A (2010) Visualization and analysis of atomistic simulation data with OVITO-the Open Visualization Tool, Modelling Simul Mater Sci Eng 18:015012.
[53] Honeycutt JD, Andersen HC (1987) Molecular-Dynamics Study of Melting and Freezing of Small Lennard-Jones Clusters, J Phys Chem-US 91:4950-4963.
[54] Frolov T, Asta M, Mishin Y (2015) Segregation-induced phase transformations in grain boundaries, Phys Rev B 92:020103.
[55] Liu XY, Xu W, Foiles SM, Adams JB (1998) Atomistic studies of segregation and diffusion in Al-Cu grain boundaries, Appl Phys Lett 72:1578-1580.
[56] Carpenter DT, Watanabe M, Barmak K, Williams DB (1999) Low-magnification quantitative X-ray mapping of grain-boundary segregation in aluminum-4 wt.% copper by analytical electron microscopy, Microsc Microanal 5:254-266.
[57] Chen Y, Gao N, Sha G, Ringer SP, Starink MJ (2016) Microstructural evolution, strengthening and thermal stability of an ultrafine-grained Al-Cu-Mg alloy, Acta Mater 109:202-212.
[58] Tsivoulas D, Robson JD (2015) Heterogeneous Zr solute segregation and $Al_3Zr$ dispersoid distributions in Al-Cu-Li alloys, Acta Mater 93:73-86.
[59] Yan HB, Gan FX, Huang DQ (1989) Evaporated Cu-Al amorphous-alloys and their phase-transition, J Non-Cryst Solids 112:221-227.
[60] Yang JJ, Yang Y, Wu K, Chang YA (2005) The formation of amorphous alloy oxides as barriers used in magnetic tunnel junctions, J Appl Phys 98:074508.
[61] Cui YY, Wang TL, Li JH, Dai Y, Liu BX (2011) Thermodynamic calculation and interatomic potential to predict the favored composition region for the Cu-Zr-Al metallic glass formation, Phys Chem Chem Phys 13:4103-4108.
[62] Daw MS, Baskes MI (1984) Embedded-atom method: Derivation and application to impurities, surfaces, and other defects in metals, Phys Rev B 29:6443.
[63] Fujita T, Guan PF, Sheng HW, Inoue A, Sakurai T, Chen MW (2010) Coupling between chemical and dynamic heterogeneities in a multicomponent bulk metallic glass, Phys Rev B 81:140204.
[64] Cheng YQ, Ma E, Sheng HW (2009) Atomic level structure in multicomponent bulk metallic glass, Phys Rev Lett 102:245501.
[65] Hu Y, Schuler JD, Rupert TJ (2018) Identifying interatomic potentials for the accurate modeling of interfacial segregation and structural transitions, Comp Mater Sci 148:10-20.
[66] Murray JL (1985) The Aluminium-Copper system. International metals reviews. 30(1):211-34.
[67] Turchanin M (1997) Calorimetric research on the heat of formation of liquid alloys of copper with group IIIA and group IVA metals, Powder Metall Met Ceram 36:253-63.
[68] Edwards RK, Downing JH (1956) The thermodynamics of the liquid solutions in the triad Cu-Ag-Au. I. The Cu-Ag system, J Phys Chem-US 60:108-111.
[69] Esin YO, Bobrov NP, Petrushevskiy MS, Geld PV (1974) Enthalpy of formation of liquid aluminum-alloys with Titanium and Zirconium, Russ Metall 5:86-89.

[70] Witusiewicz VT, Hecht U, Fries SG, Rex S (2004) The Ag-Al-Cu system: Part I: Reassessment of the constituent binaries on the basis of new experimental data, J Alloys Comp 385:133-143.
[71] Jain A, Ong SP, Hautier G, Chen W, Richards WD, Dacek S, Cholia S, Gunter D, Skinner D, Ceder G (2013) Commentary: The Materials Project: A materials genome approach to accelerating materials innovation, APL Mater 1:011002.
[72] Lazarus D (1949) The variation of the adiabatic elastic constants of KCl, NaCl, CuZn, Cu, and Al with pressure to 10,000 bars, Phys Rev 76:545.
[73] Hearmon RFS (1946) The elastic constants of anisotropic materials, Rev Mod Phys 18:409.
[74] Hearmon RFS (1956) The elastic constants of anisotropic materials—II. Advances in Physics, 5:323-382.
[75] Straumanis ME, Yu LS (1969) Lattice parameters, densities, expansion coefficients and perfection of structure of Cu and of Cu-in alpha phase, Acta Cryst 25:676-682.
[76] Hertzberg RW (1996) Deformation and fracture and fracture mechanics of engineering materials 4th edn. Wiley, New York.
[77] Methfessel M, Hennig D, Scheffler M (1992) Trends of the surface relaxations, surface energies, and work-functions of the 4d transition-metals, Phys Rev B 46:4816-4829.
[78] Liu LG, Bassett WA (1973) Compression of Ag and phase transformation of NaCl, J Appl Phys 44:1475-9.
[79] Straumanis ME, Woodward CL (1971) Lattice parameters and thermal expansion coefficients of Al, Ag and Mo at low temperatures. Comparison with dilatometric data, Acta Cryst 27:549-551.
[80] Yang S, Zhou N, Zheng H, Ong SP, Luo J (2018) First-Order Interfacial Transformations with a Critical Point: Breaking the Symmetry at a Symmetric Tilt Grain Boundary, Phys Rev Lett 120:085702.
[81] Tewari A, Galmarini S, Stuer M, Bowen P (2012) Atomistic modeling of the effect of codoping on the atomistic structure of interfaces in alpha-alumina, J Eur Ceram Soc 32:2935-2948.
[82] Huang ZF, Chen F, Shen Q, Zhang L, Rupert TJ, work in preparation. Combined effects of nonmetallic impurities and planned metallic dopants on grain boundary energy and strength.
[83] Dieter GE (1986) Mechanical metallurgy 3$^{rd}$ edn, McGraw-Hill, New York.
[84] Chen N, Niu LL, Zhang Y, Shu X, Zhou HB, Jin S, Ran G, Lu GH, Gao F (2016) Energetics of vacancy segregation to [100] symmetric tilt grain boundaries in bcc tungsten. Scientific reports, 6:36955.
[85] Zhou X, Song J (2017) Effect of local stress on hydrogen segregation at grain boundaries in metals, Mater Lett 196:123-7.
[86] Liu XY, Adams JB (1998) Grain-boundary segregation in Al-10% Mg alloys at hot working temperatures, Acta Mater 46:3467-3476.
[87] Wang D, Tan H, Li Y (2005) Multiple maxima of GFA in three adjacent eutectics in Zr–Cu–Al alloy system − A metallographic way to pinpoint the best glass forming alloys, Acta Mater 53: 2969-2979.
[88] Wang XD, Jiang QK, Cao QP, Bednarcik J, Franz H, Jiang JZ (2008) Atomic structure and glass forming ability of $Cu_{46}Zr_{46}Al_8$ bulk metallic glass, J Appl Phys 104:093519.
[89] Inoue A, Zhang W (2002) Formation, thermal stability and mechanical properties of Cu-Zr-Al bulk glassy alloys, Mater Trans 43:2921-2925.

**Table 1** Physical properties of Cu, Al, Ag, and Zr$_{fcc}$ from experiments/ab initio calculations and reproduced by the interatomic potentials used in the present study.

| Interatomic potential | Element type | Bulk modulus (GPa) | | Shear modulus, C$_{44}$ (GPa) | | Surface energy (1 0 0) (J/m²) | | Surface energy (1 1 1) (J/m²) | | Lattice parameter (Å) | |
|---|---|---|---|---|---|---|---|---|---|---|---|
| | | Present work | Reported data | Present work | Reported data | Present work | Reported data | Present work | Reported data | Present work | Reported data |
| Cu-Zr-Ag | Cu | 150.47 | 139[71], 143.37[72] | 73.15 | 79[71], 76.10[72], 76.27[72], 75.4[76] | 1.74 | 1.47[71] | 1.48 | 1.31[71] | 3.601 | 3.616[71], 3.615[75] |
| | Zr$_{fcc}$ | 85.43 | 90[71] | | | 1.43 | 1.42[71] | | | 4.528 | 4.537[71] |
| | Ag | 99.77 | 88[71], 99.80[73], 103.60[74] | 47.67 | 43.67[73], 46.10[74] | 0.97 | 0.81[71], 1.21[77] | 0.90 | 0.77[71], 1.21[77] | 4.068 | 4.161[71], 4.085[78] |
| Al-Zr-Cu | Al | 82.98 | | 35.26 | 32[71], 30.32[72], 30.82[72], 28.5[76] | 0.89 | 0.92[71] | 0.85 | 0.80[71] | 4.017 | 4.039[71], 4.035[79] |
| | Zr$_{fcc}$ | 96.30 | 90[71] | | | 1.41 | 1.42[71] | | | 4.528 | 4.537[71] |
| | Cu | 135.77 | 139[71], 143.37[72] | 73.47 | 79[71], 76.10[72], 76.27[72], 75.4[73] | 1.68 | 1.47[71] | 1.46 | 1.31[71] | 3.598 | 3.616[71], 3.615[75] |

Note: $Zr_{fcc}$ refers to Zr face-centered-cubic crystal structure. Data from Ref. [71] was obtained by density functional theory (DFT) calculation using the projector augmented wave approach and the Perdew-Burke-Ernzerhof exchange-correlation GGA functional, and was stored in the database of the Materials Project (www.materialsproject.org). The elastic constants in Ref. [72], [73], and [74] were measured using dynamic methods and the bulk modulus was then calculated using ⅓($C_{11}$+2$C_{12}$). The lattice constant of Cu from Ref. [75] was obtained experimentally at room temperature using the asymmetric film method. Data from Ref. [77] was also obtained by DFT calculations but with a full-potential linear-muffin-tin-orbital method with some modifications to better describe surfaces. The lattice parameter of Ag from Ref. [78] was measured by X-ray diffraction at room temperature, and the lattice parameter of Al from Ref. [79] was measured by X-ray diffraction at 125 K.

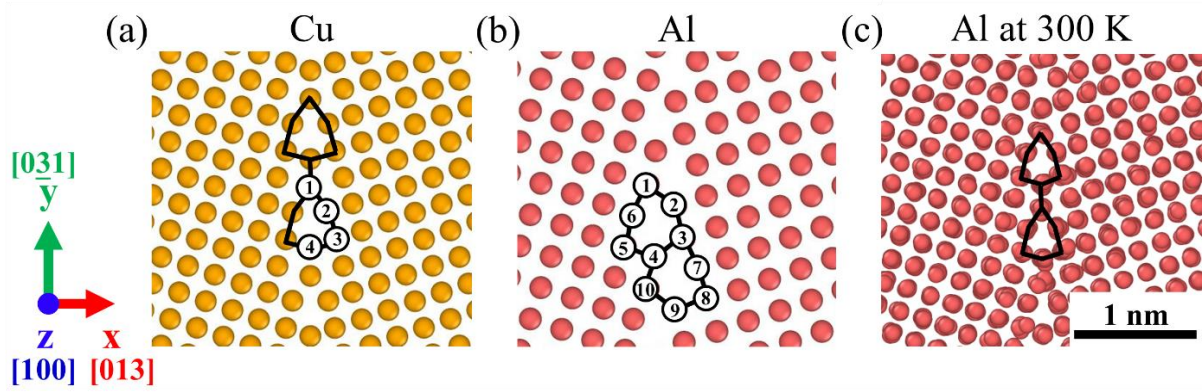

**Figure 1** A Σ5 (310) grain boundary in **a** pure Cu and **b** pure Al (i.e. a clean grain boundary) at 0 K. **c** The same boundary in pure Al at 300 K. The repeating kite-shape structure unit is outlined by black lines and certain sites are numbered for the calculation of segregation energy of dopants to different positions in the grain boundary.

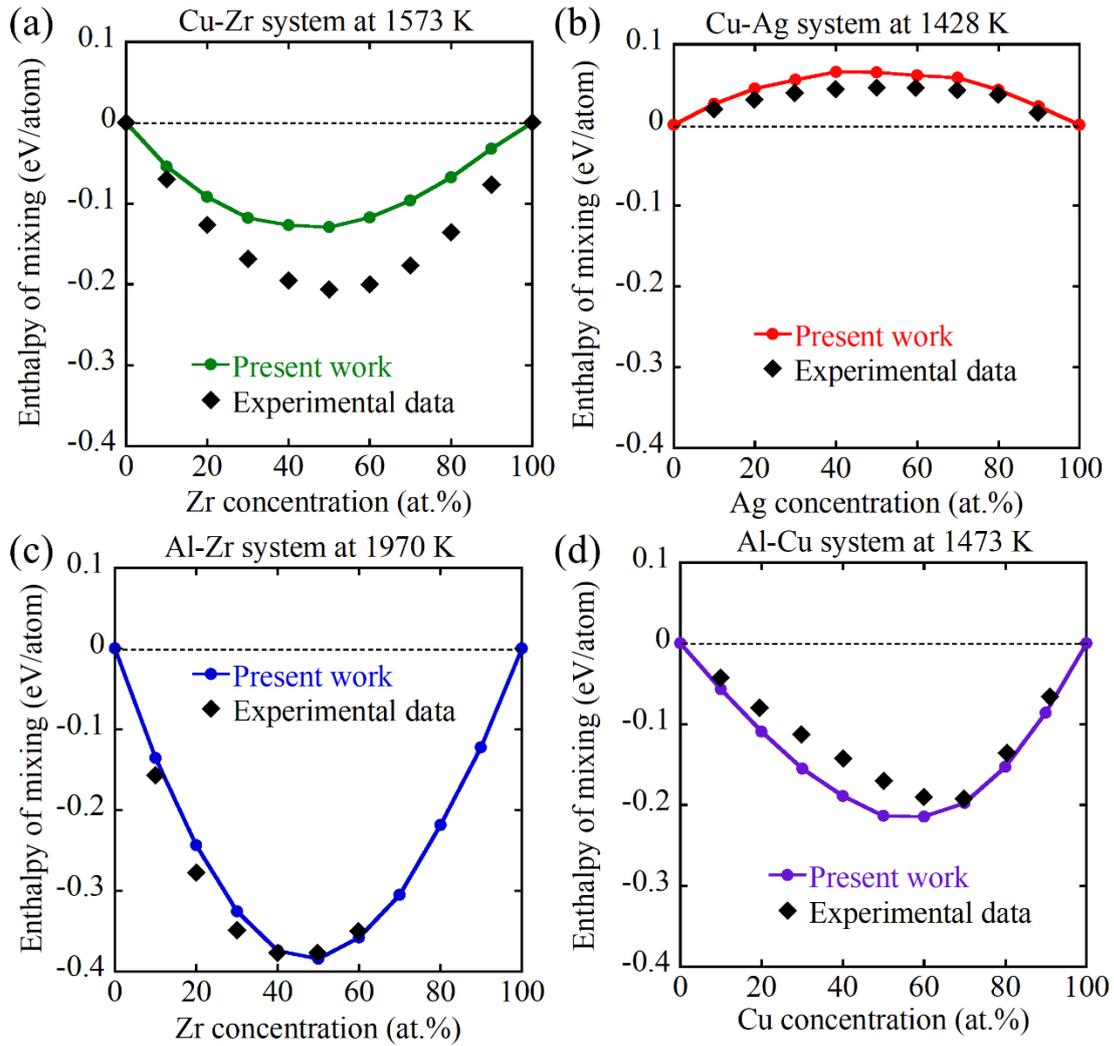

**Figure 2** Enthalpy of mixing of **a** the Cu-Zr system at 1573 K, **b** the Cu-Ag system at 1428 K, **c** the Al-Zr system at 1970 K, and **d** the Al-Cu system at 1473 K. The experimental data comes from **a** Turchanin et al. [67], **b** Edwards et al. [68], **c** Esin et al. [69], and **d** Witusiewicz et al. [70].

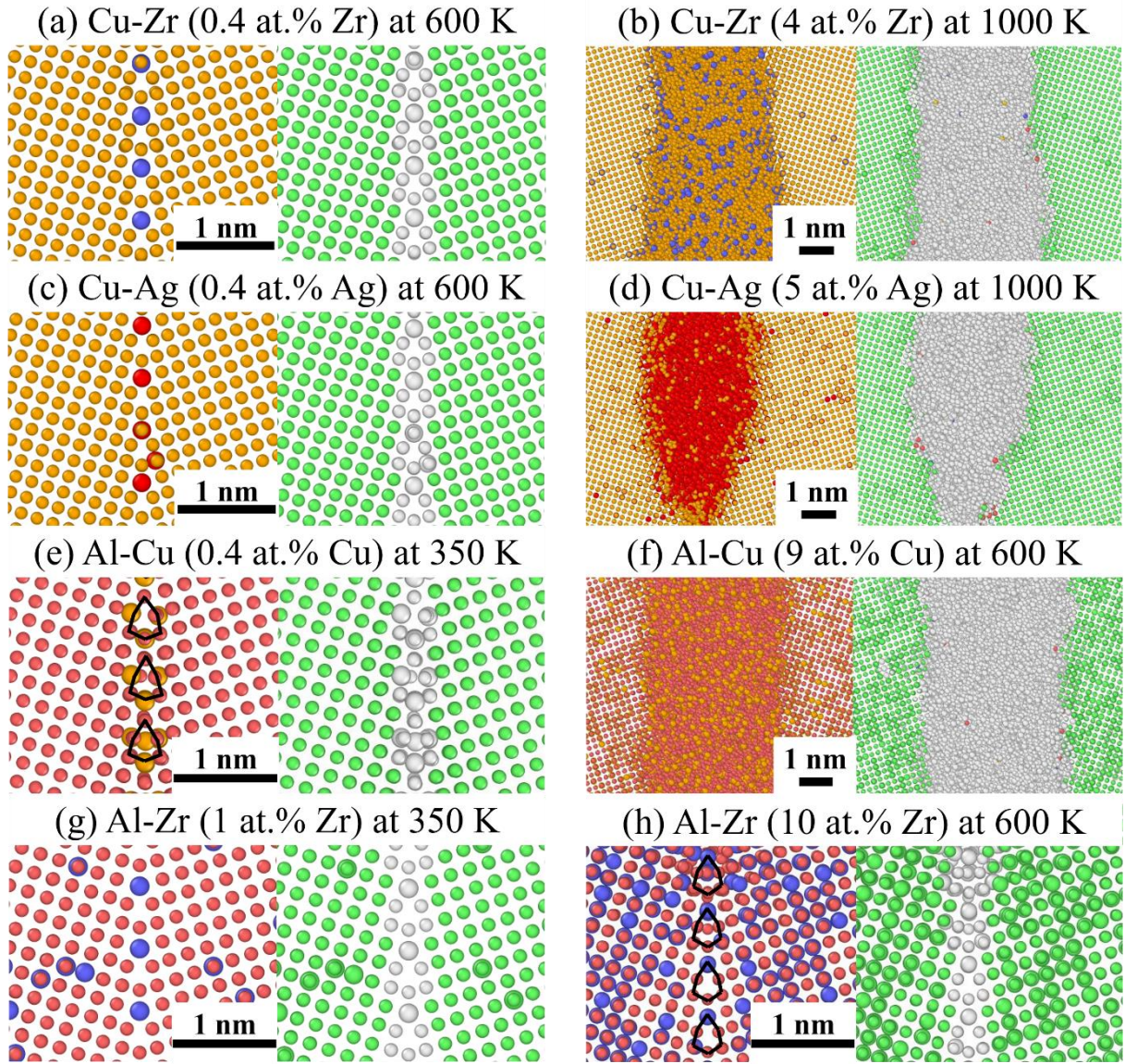

**Figure 3** The chemical (left frame) and structural (right frame) information of the Σ5 (013) grain boundary in Cu doped with **a** 0.4 at.% Zr at 600 K, **b** 4 at.% Zr at 1000 K, **c** 0.4 at.% Ag at 600 K, and **d** 5 at.% Zr at 1000 K, as well as the Σ5 (013) grain boundary in Al doped with **e** 0.4 at.% Cu at 350 K, **f** 9 at.% Cu at 600 K, **g** 1 at.% Zr at 350 K, and **h** 10 at.% Zr at 600 K. In the left panels, Cu atoms are colored orange, Ag atoms are colored red, Zr atoms are colored blue, and Al atoms are colored pink. In the right panels, face centered cubic atoms are colored green, hexagonal close packed atoms are colored red, body centered cubic atoms are colored blue, icosahedral atoms are colored yellow, and "other" atoms are colored white. The atomic radii of dopants are set slightly larger than that of matrix element to show them more clearly. The kite-shape structure at the grain boundary is outlined by black lines in **e** and **h**.

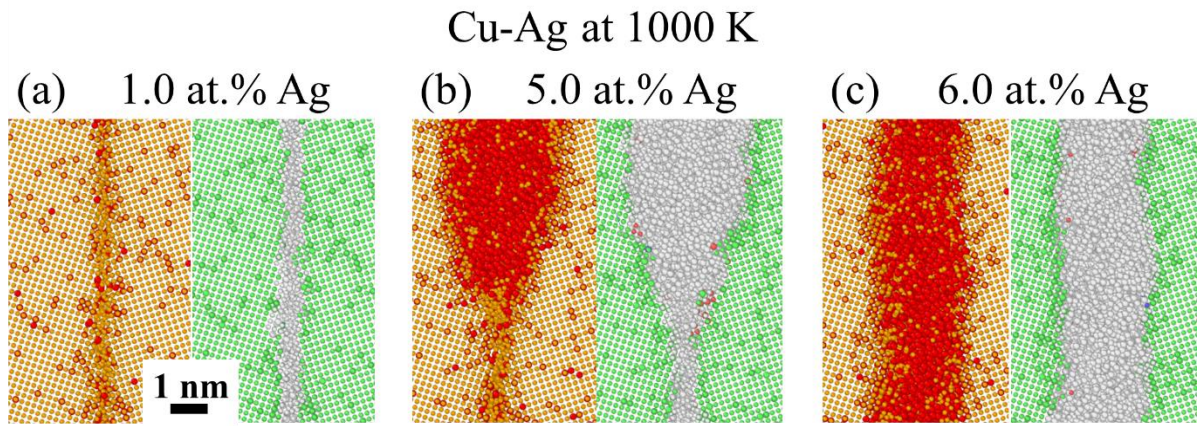

**Figure 4** The chemical (left frame) and structural (right frame) information of the Σ5 (013) grain boundary in Cu that has been doped with **a** 1 at.% Ag, **b** 5 at.% Ag, and **c** 6 at.% Ag at 1000 K. In the left panels, Cu atoms are colored orange and Ag atoms red. In the right panels, face centered cubic atoms are colored green, hexagonal close packed atoms are colored red, body centered cubic atoms are colored purple, icosahedral atoms are colored yellow, and "other" atoms are colored white. The atomic radii of dopants are set slightly larger than that of matrix element to show them more clearly.

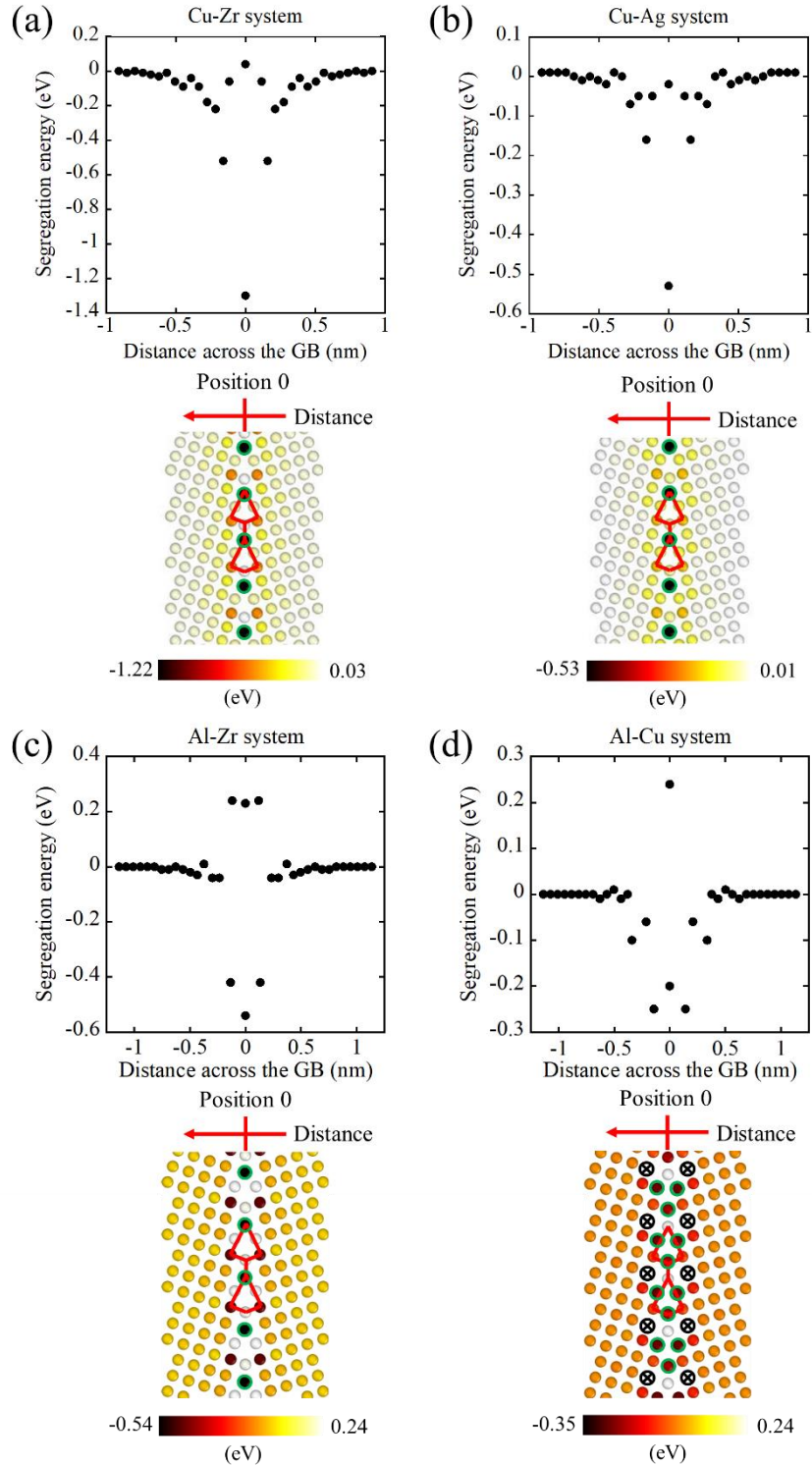

**Figure 5** The spatial dependence of the segregation energy of **a** Zr in Cu, **b** Ag in Cu, **c** Zr in Al and **d** Cu in Al across the grain boundary (top frames). The images below each figure provide a reference for how distance across the grain boundary is measured. A positive value of segregation energy indicates depletion from the grain boundary, while a negative value indicates segregation of dopant to the grain boundary. In the bottom frames, the kite-shape structure at the grain boundary is outlined by red lines, and the promising segregation sites of each solute is outlined by green circles. In **d**, data for the atoms with cross signs on them is not shown, since the grain boundary already becomes asymmetric at 100 minimization step.

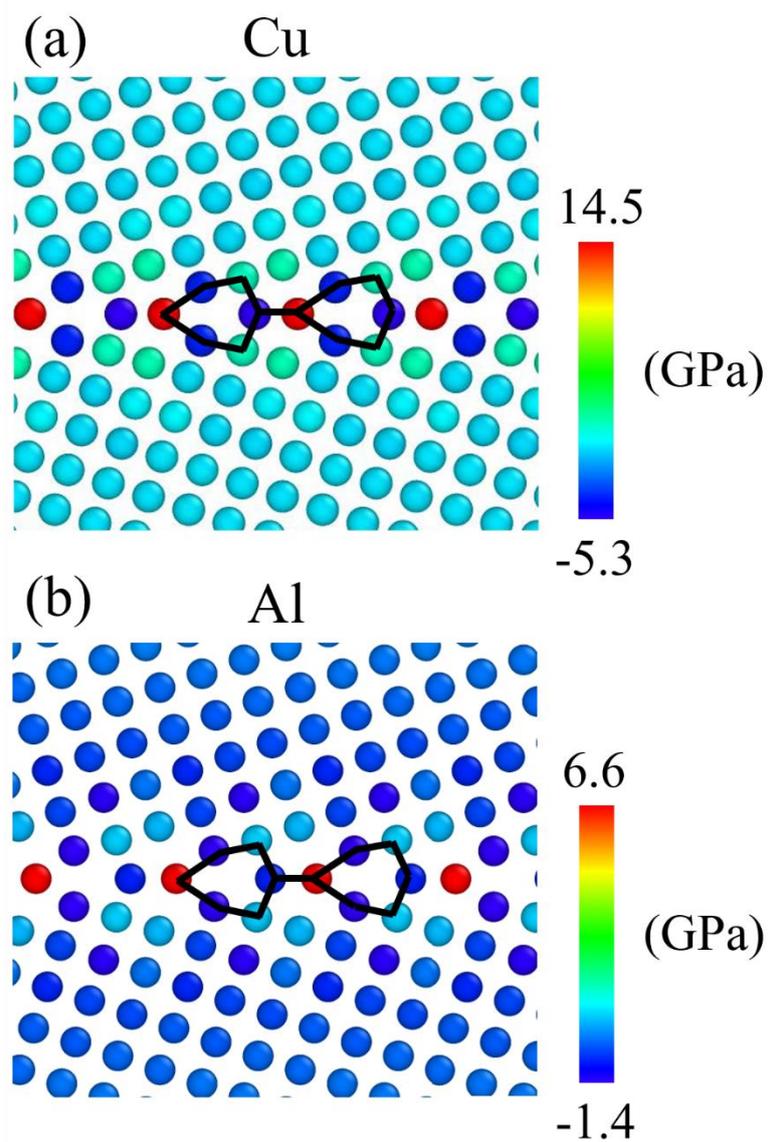

**Figure 6** The distribution of atomic hydrostatic stress across the Σ5 (310) grain boundary in **a** pure Cu and **b** pure Al. Positive values indicate sites under tension, while negative values indicate sites under compression. The kite-shape structure at the grain boundary is outlined by black lines.

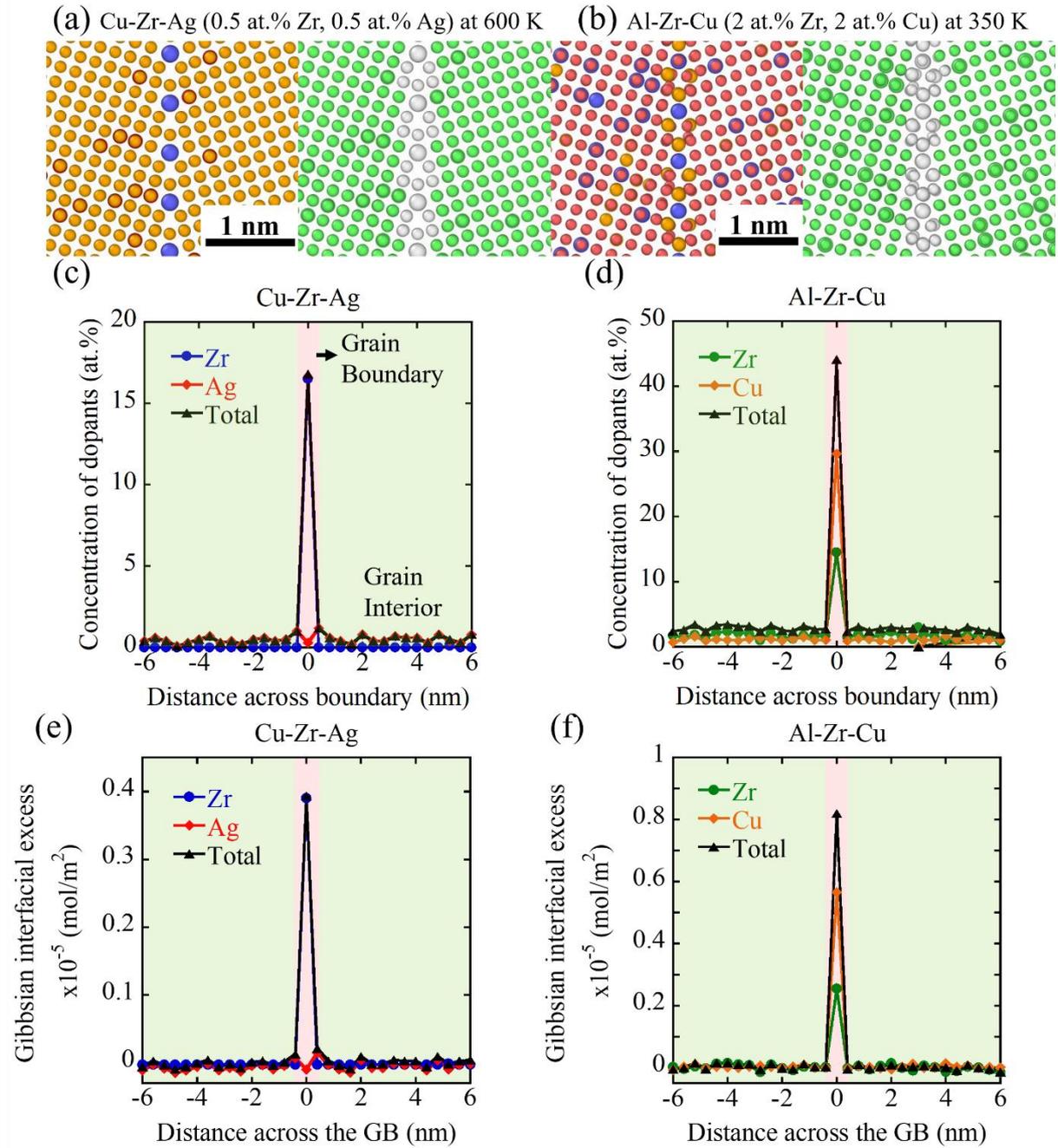

**Figure 7** The chemical (left frame) and structural (right frame) information of the Σ5 (013) grain boundary in Cu doped with **a** 0.5 at.% Zr and 0.5 at.% Ag at 600 K, **b** 2 at.% Zr and 2 at.% Cu at 350 K. In the left panels, Cu atoms are colored orange, Ag atoms are colored red, Zr atoms are colored blue, and Al atoms are colored pink. In the right panels, face centered cubic atoms are colored green and other atoms are colored white. The concentration profile of dopants across the grain boundary for **c** Cu-0.5 at.% Zr-0.5 at.% Ag and **d** Al-2 at.% Zr-2 at.% Cu. The Gibbsian interfacial excess profile of solutes across the grain boundary for **e** Cu-0.5 at.% Zr-0.5 at.% Ag and **f** Al-2 at.% Zr-2 at.% Cu. In **c**, **d**, **e** and **f**, the light green regions are the grain interiors and the light red regions are the grain boundaries.

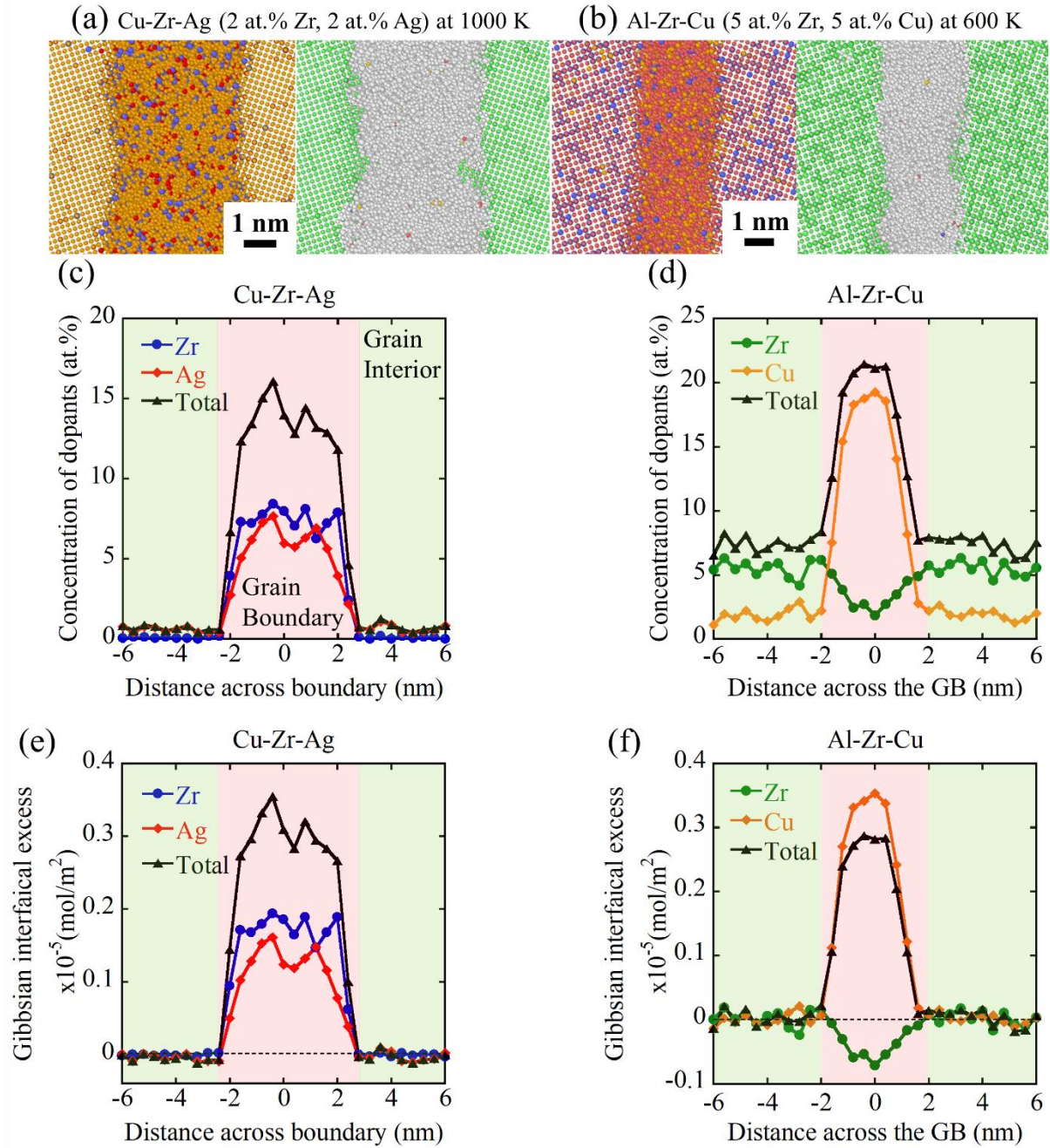

**Figure 8** The chemical (left frame) and structural (right frame) information of the Σ5 (013) grain boundary in Cu doped with **a** 2 at.% Zr and 2 at.% Ag at 1000 K, **b** 5 at.% Zr and 5 at.% Cu at 600 K. In the left panels, Cu atoms are colored orange, Ag atoms are colored red, Zr atoms are colored blue, and Al atoms are colored pink. In the right panels, face centered cubic atoms are colored green, hexagonal close packed atoms are colored red, body centered cubic atoms are colored purple, icosahedral atoms are colored yellow, and other atoms are colored white. The concentration profile of dopants across the grain boundary for **c** Cu-2 at.% Zr-2 at.% Ag and **d** Al-5 at.% Zr-5 at.% Cu. The Gibbsian interfacial excess profile of dopants across the grain boundary for **e** Cu-2 at.% Zr-2 at.% Ag and **f** Al-5 at.% Zr-5 at.% Cu. In **c**, **d**, **e** and **f**, the light green regions are the grain interiors and the light red regions are the grain boundaries.

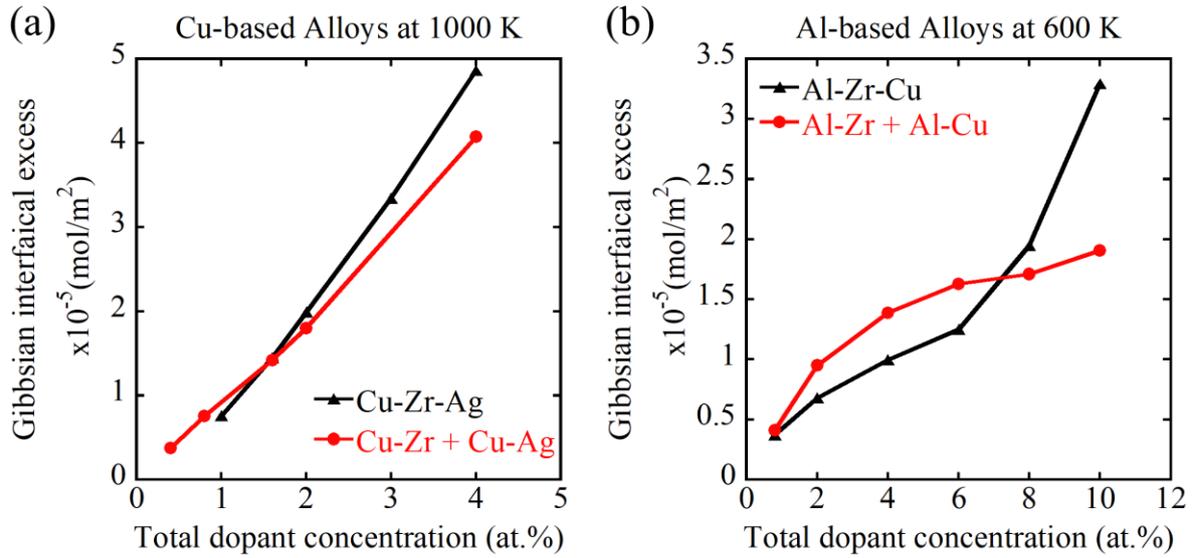

**Figure 9** The Gibbsian interfacial excess of dopants as a function of the total dopant concentration for **a** Cu-Zr-Ag ternary alloys with a 1:1 ratio of Zr and Cu versus a summation of the Cu-Zr + Cu-Ag binary behaviors, and for **b** Al-Zr-Cu ternary alloys with a 1:1 ratio of Zr and Cu versus a summation of Al-Zr + Al-Cu binary behaviors. In **a**, AIFs start to form at ~1.5 at.% of dopants where the two curves intersect. While in **b**, AIFs form before the two curves intersect at ~7.5 at.%.

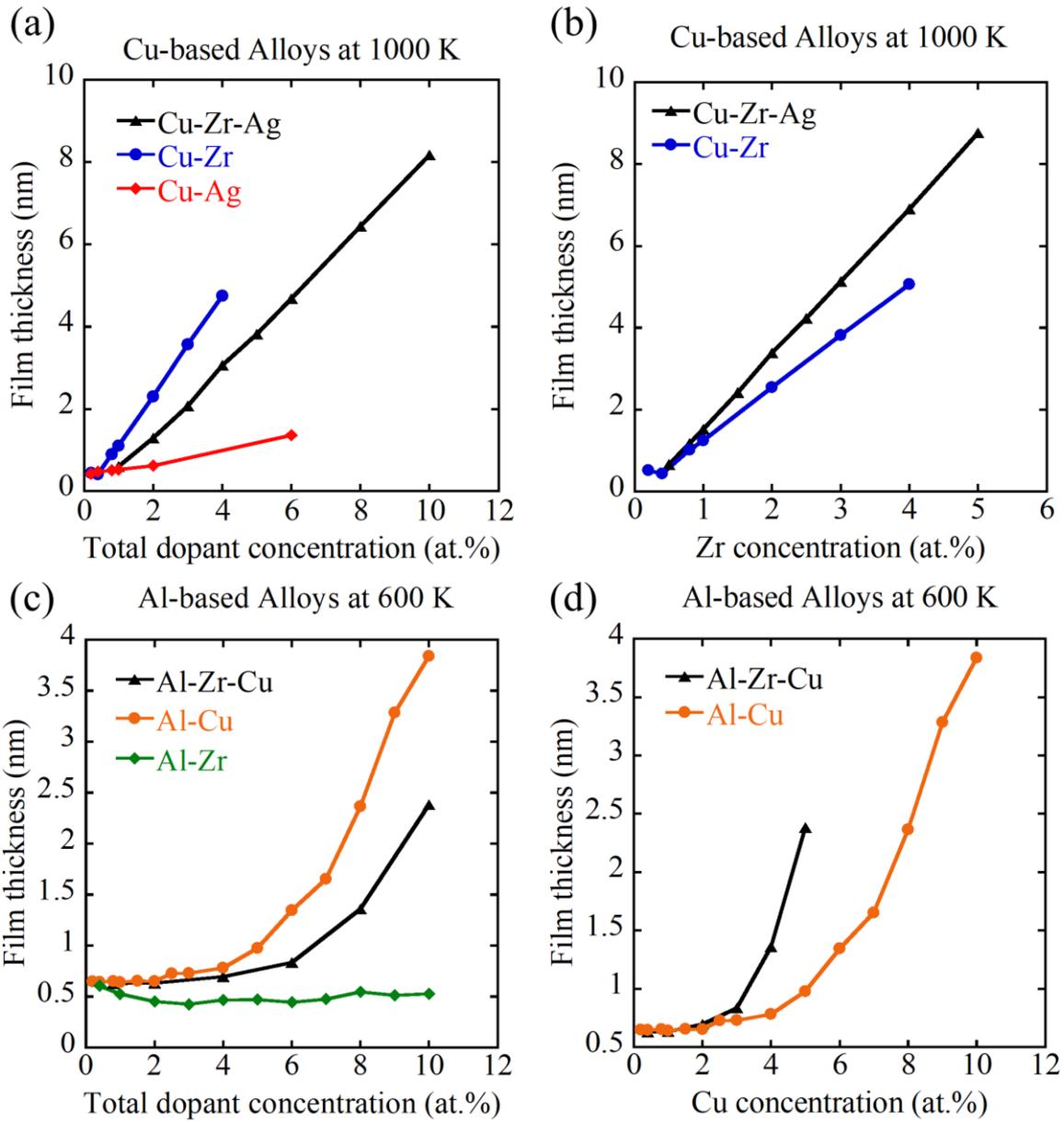

**Figure 10** The variation of film thickness with increasing total dopant concentration for **a** Cu-Zr, Cu-Ag, and Cu-Zr-Ag with a 1:1 ratio of Zr and Ag, as well as film thickness as a function of Zr concentration for **b** Cu-Zr and Cu-Zr-Ag. The variation of film thickness with increasing total dopant concentration for **c** Al-Zr, Al-Cu, and Al-Zr-Cu with a 1:1 ratio of Zr and Cu, as well as film thickness as a function of Cu concentration for **d** Al-Cu and Al-Zr-Cu.

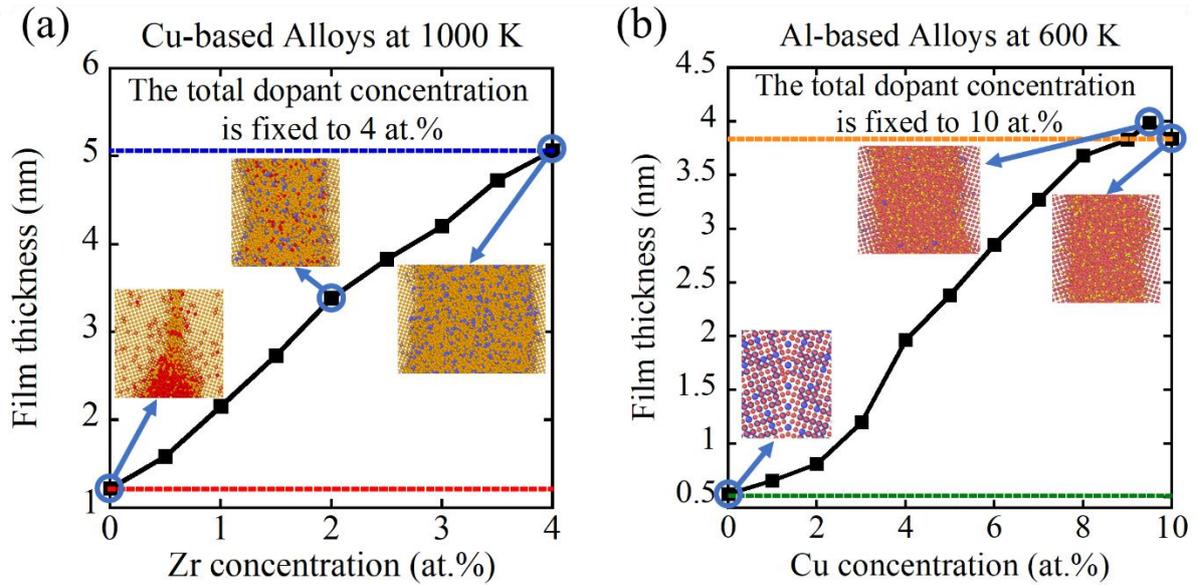

**Figure 11 a** The film thickness as a function of Ag concentration for Cu-Zr-Ag. The total dopant concentration is fixed as 4 at.%, meaning the system with 0 at.% Ag is the Cu-Zr binary (blue dashed line) while the system with 4 at.% Ag is the Cu-Ag binary (red dashed line). The insets show the atomic structures of the grain boundary in the Cu-Zr, Cu-Ag, and Cu-2 at.% Zr-2 at.% Ag. **b** The film thickness as a function of Cu concentration for Al-Zr-Cu alloys. The total dopant concentration is fixed as 10 at.%, meaning the system with 0 at.% Cu is the Al-Zr binary (green dashed line) while the system with 10 at.% Cu is the Al-Cu binary (orange dashed line). The insets show the atomic structures of the grain boundary in the Al-Zr, Al-Cu, and Al-0.5 at.% Zr-9.5 at.% Cu.